\documentclass[
    ,final            
  ,numberedheadings 
  ]
  {aipproc}

\layoutstyle{8x11single}

\begin{document}

\begin{flushright}
  FERMILAB-CONF-08-050-T
\end{flushright}
\vspace*{-0.5in}

\title{Differentiating Neutrino Models on the Basis of $\theta_{13}$
      and Lepton Flavor Violation\footnote{Talk presented at the 
      International Workshop on Grand Unified Theories: Current Status
      and Future Prospects (GUT07), December 17-19, 2007, Kusatsu, Japan}}

\author{Carl H. Albright\footnote{email: albright@fnal.gov}}{
  address={Department of Physics, Northern Illinois University, DeKalb, 
      Illinois 60115, USA}
  ,altaddress={Fermi National Accelerator Laboratory, Batavia, Illinois 
      60510, USA}
}


\begin{abstract}
 
 We show how models of neutrino masses and mixings can be differentiated
 on the basis of their predictions for $\theta_{13}$ and lepton flavor violation
 in radiative charged lepton decays and $\mu - e$ conversion.  We illustrate
 the lepton flavor violation results for five predictive $SO(10)$ SUSY GUT
 models and point out the relative importance of their heavy right-handed
 neutrino mass spectra and $\theta_{13}$ predictions.
 
\end{abstract}

\maketitle

\vspace*{-0.25in}
\section{Introduction}

Many models exist in the literature which attempt to explain the observed
neutrino masses and mixings.  The viable models agree with the presently
known neutrino oscillation parameters falling within the $2\sigma$ ranges
\cite{Maltoni:2004ei}:
\begin{equation}
\begin{tabular}{rclrcl}
   $\sin^2 \theta_{12}$ &=& 0.28 - 0.37, \qquad &   $\Delta m^2_{21}$ &=& 
   	$(7.3 - 8.1) \times 10^{-5}\ {\rm eV^2}$,\\
    $\sin^2 \theta_{13}$ &=& 0.38 - 0.63,  &     $\Delta m^2_{31}$ &=& 
    	$(2.0 - 2.8) \times 10^{-3}\ {\rm eV^2}$,\\
   $\sin^2 \theta_{13}$ &$\leq$& 0.033.\\ 
\end{tabular}
\label{eq:data}
\end{equation}
\noindent The data suggests the approximate tri-bimaximal 
mixing texture of Harrison, Perkins, and Scott \cite{tbm}:
\begin{equation}
  U_{PMNS} = \left( \begin{array}{ccc} 2/\sqrt{6} & 1/\sqrt{3} & 0\\
             -1/\sqrt{6} & 1/\sqrt{3} & 1/\sqrt{2}\\
	   -1/\sqrt{6} & 1/\sqrt{3} & 1/\sqrt{2} \end{array}\right),
\label{eq:PMNS}
\end{equation}

\noindent with $\sin^2 \theta_{23} = 0.5,\ \sin^2 \theta_{12} = 0.33$,
and $\sin^2 \theta_{13} = 0$.

The reason for the plethera of models still in agreement with experiment
of course can be traced to the inaccuracy of the present data.  In 
addition, there are a number of unknowns that must be still determined:  
the hierarchy and absolute mass scales of the light neutrinos; the Dirac 
or Majorana nature of the neutrinos; the CP-violating phases of the mixing 
matrix; any departure of the reactor neutrino angle, $\theta_{13}$
from $0^\circ$;
any departure of the atmospheric neutrino mixing angle, $\theta_{23}$, 
from $45^\circ$; whether the approximate tri-bimaximal mixing is a 
softly-broken or accidental symmmetry; and the magnitude of the charged 
lepton flavor violation?  In this presentation we survey some of the models 
to determine what they predict for $\theta_{13}$, the hierarchy, and 
lepton flavor violation.

\section{Theoretical framework and models}

The observation of neutrino oscillations implies that neutrinos have mass,
with the mass squared differences given in Eq.(\ref{eq:data}).  Information
concerning the absolute neutrino mass scale has been determined by the 
combined WMAP, SDSS, and Lyman alpha data \cite{WMAP} which place an upper 
limit on the sum of the masses, $\sum_i m_i \leq 0.17$ eV.
An extension of the SM is then required, and possible suggestions
include one or more of the following:
\begin{itemize}
\item the introduction of dim-5 effective non-renormalizable operators;
\item the addition of right-handed neutrinos with their Yukawa couplings
  to the left-handed neutrinos;
\item the addition of direct mass terms with right-handed Majorana couplings;
\item the addition of a Higgs triplet with left-handed Majorana couplings.
\end{itemize}

The general $6 \times 6$ neutrino mass matrix in the $B(\nu_{\alpha L},\ 
N^c_{\alpha L})$ flavor basis of the six left-handed fields then has the 
following structure in terms of $3 \times 3$ submatrices:
\begin{equation}
  \cal{M} = \left(\begin{array}{cc} M_L & M^T_N \\ M_N & M_R \end{array}
             \right),
\label{eq:6x6}
\end{equation}

\noindent where $M_N$ is the Dirac neutrino mass matrix, $M_L$ the 
left-handed and $M_R$ the right-handed Majorana neutrino mass matrices.
With $M_L = 0$ and $M_N << M_R$, the type I seesaw formula is obtained
\cite{seesaw} 
\begin{equation}
  M_\nu = - M^T_N M^{-1}_R M_N,
\label{eq:typeI}
\end{equation}

\noindent for the light Majorana neutrinos, while if $M_L \neq 0$ and 
$M_N << M_R$, one obtains the type II seesaw formula \cite{seesawII},
\begin{equation}
  M_\nu = M_L - M^T_N M^{-1}_R M_N.
\label{eq:typeII}
\end{equation}

The effective light Majorana mass matrix is complex symmetric
and can be diagonalized by a unitary transformation, $U_{\nu L}$, to give
\begin{equation}
  M^{diag}_\nu = U^T_{\nu L} M_\nu U_{\nu L} = {\rm diag}(m_1,\ m_2,\ m_3),
\end{equation}

\noindent with real, positive masses down the diagonal.  On the other hand,
the Dirac charged lepton mass matrix is diagonalized by a bi-unitary 
transformation according to 
\begin{equation}
  M^{diag}_E = U^\dagger _{ER} M_EU_{EL} = {\rm diag}(m_e,\ m_\mu, m_\tau).
\end{equation}
The neutrino mixing matrix, $V_{PMNS}$ \cite{PMNS}, is then given by \\
\begin{equation}
\begin{array}{rcl}
  V_{PMNS}&\equiv&U^\dagger _{EL} U_{\nu L} = U_{PMNS}\Phi,\\[0.1in]
  \Phi&=&{\rm diag}(e^{i\chi_1},\ e^{i\chi_2},1),\\
\end{array}
\label{eq:VPMNS}
\end{equation}

\noindent in terms of the approximately tri-bimaximal mixing matrix,
$U_{PMNS}$ and the phase matrix, $\Phi$, since an arbitrary phase rotation
of $U_{\nu L}$ is not possible in the above.

Models which have been introduced to explain the neutrino oscillation 
phenomena fall into two categories.  There are those based on some 
lepton flavor symmetry such as $\mu - \tau$ interchange symmetry, the more
restrictive $S_3$ or $A_4$ flavor symmetry, and $SO(3)$ or $SU(3)$ flavor
symmetries.  Many attempts have been made to explore the location of 
texture zeros for the lepton mass matrices, with the hope that some flavor
symmetry might be identified that way.  A more ambitious class of models is 
based on $SU(5)$ or $SO(10)$ grand unification, where one attempts to 
explain the masses and mixings in the quark sector as well as the lepton 
sector.  These models are said to have a ``minimal'' Higgs structure, if 
the Higgs bosons responsible for electroweak symmetry breaking transform 
as the ${\bf 10,\ 126}$ dimensional representations of $SO(10)$, and 
possibly the ${\bf 120,\ 45}$ or ${\bf 54}$ representations.  They lead to
symmetric or antisymmetric entries in the mass matrices.  Other $SO(10)$ 
models have Higgs bosons which transform as ${\bf 10,\ 16}$, $\overline{16}$,
and ${\bf 45}$ representations and are referred to as ``lopsided,''  
since lopsided contributions to the down quark and charged lepton
mass matrices can occur due to the $SU(5)$ structure of the ${\bf 16}$'s.
For recent reviews, cf. Ref. \cite{Chen:2003zv}.

\section{Survey of Model Predictions for $\theta_{13}$}

In a previous publication \cite{Albright:2006cw} we have made a survey of 
63 models in the literature which give the large mixing angle (LMA) solution 
for the solar neutrino oscillations and firm and reasonably restrictive 
predictions for the reactor neutrino angle.  The cutoff date for the selection 
of models was chosen to be May 2006.  In this study we found most of the 
models predict a value in the range of $10^{-4} < \sin^2 \theta_{13} < 0.04$ 
with a normal hierarchy preferred by 3:1.  The results are displayed in 
Fig. 1 in the form of a $\sin^2 \theta_{13}$ histogram, where the models are 
distinguished according to their type, and each is assigned the same area on 
the histogram though their imprecise predictions may cover several bins.

\begin{figure}
\includegraphics*[scale=0.6]{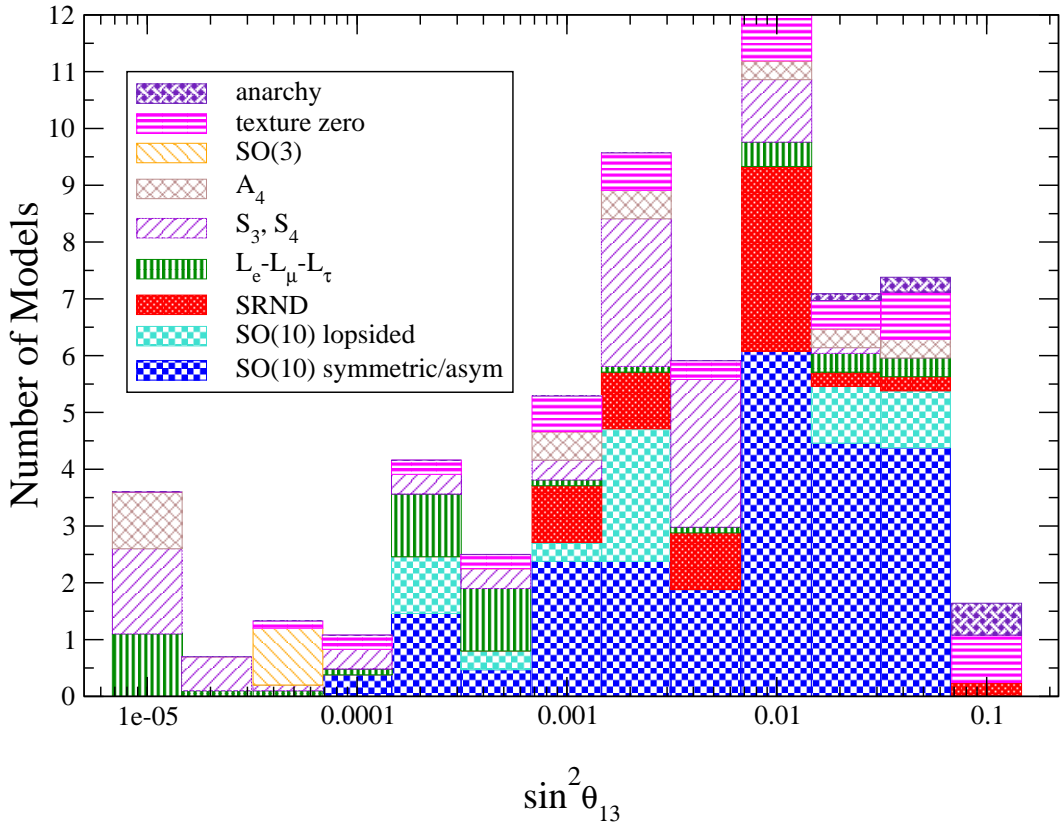}
\caption{Histogram of $\sin^2 \theta_{13}$ predictions for 
  all 63 models.}
\end{figure}

Since the planned Double CHOOZ and Daya Bay reactor experiments 
\cite{newreactors} will reach $\sin^2 2\theta_{13} \sim 
0.01$, roughly half of the models will be eliminated, either by the 
positive or negative signal of disappearance of $\bar{\nu}_e$'s from their 
beams.   But even if a positive signal for the disappearance is seen,  
Fig. 1. indicates that of the order of 5 - 10 models will still survive,
depending upon the accuracy of the measurement and the resilience of the 
models.  

Meanwhile the MEG experiment \cite{MEG} at PSI is beginning to look for the
$\mu \rightarrow e \gamma$ decay mode.  With plans to lower the present 
branching ratio limit \cite{MEGA} of $1.2 \times 10^{-11}$ down to the 
$10^{-13}$ range, this experiment may serve as an even more immediate 
selector of models.  With this in mind, we turn to the subject of charged 
lepton flavor violation as a further distinguishing feature of the models 
proposed.

\section{Lepton Flavor Violation in Radiative Decays}

\begin{figure}[b]
\includegraphics*[scale=0.8]{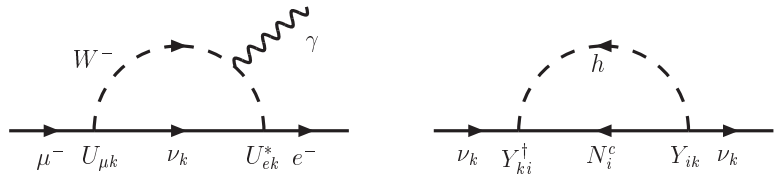}\\[-0.1in]
\caption{Example of a Feynman diagram for $\mu \rightarrow e 
\gamma$ with a neutrino mass insertion in the SM.}
\end{figure}

It is of interest to look first at charged lepton flavor violation in the
SM with the addition of three massive right-handed neutrinos.  In this 
case, individual $L_e,\ L_\mu$ and $L_\tau$ lepton numbers are not 
conserved since Majorana mass terms are not forbidden.  The lepton flavor
violation then arises in 1-loop diagrams, where the neutrino insertion
involves lepton flavor-changing Yukawa couplings.  Examples of such 
diagrams are given in Fig. 2 for the process $\mu \rightarrow e \gamma$.
\noindent The $U_{\ell k}$ are elements of the PMNS mixing matrix, while 
the $Y_{ik}$ are Yukawa couplings for $i,k = 1,2,3$, and $N^c_i$ is one of 
the heavy left-handed conjugate neutrinos.  The branching ratio is given
by \cite{LeeShrock}
\begin{equation}
\begin{array}{rcl}
  BR21&\equiv&\Gamma(\mu \rightarrow e\gamma)/\Gamma(\mu \rightarrow \nu_\mu
         e\bar{\nu}_e)\\
         &=& \frac{3\alpha}{32\pi}\left|\sum_k U^*_{\mu k}\frac{m^2_k}{M^2_W}
			U_{ke}\right|^2\\
         &\simeq& {3\alpha}{128\pi}\left(\frac{\Delta m^2_{21}}{M^2_W}\right)^2
		  	\sin^2 2\theta_{12} \sim 10^{-54}. \\
\end{array}
\label{eq:BR21SM}			
\end{equation}

\noindent In this extended SM, the branching ratio is immeasurably
small, due to the approximate GIM cancellation and the extremely small
mass ratios of the left-handed neutrinos to the massive right-handed 
neutrinos.  For such models, the MEG experiment would be expected to give
a null result.
 
In SUSY GUT models, slepton - neutralino and sneutrino - chargino loops
contribute to the radiative decays as shown in Fig. 3.   

\begin{figure}[h]
\includegraphics*[scale=0.8]{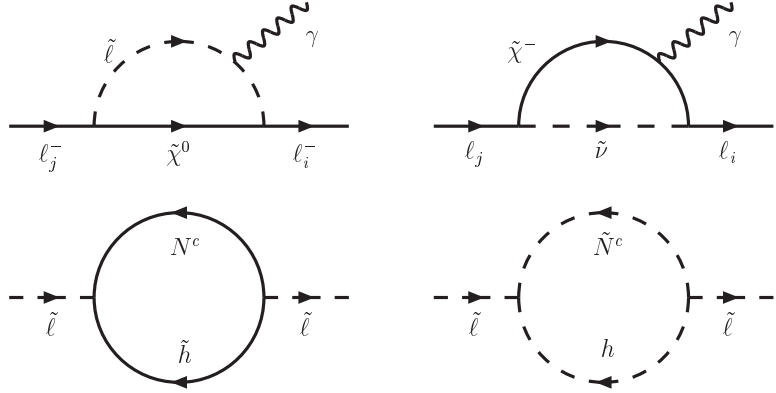}\\[-0.1in]
\caption{Examples of Feynman diagrams for slepton - neutralino 
 and sneutrino - chargino contributions to $\mu \rightarrow e \gamma$ in 
 SUSY models with slepton mass insertions.}
 \end{figure}

\noindent In the constrained MSSM (CMSSM) with universal soft masses and 
trilinear couplings \cite{CMSSM}, charged lepton flavor violation arises 
from evolution of the Yukawa couplings and soft SUSY-breaking parameters.  
Thus with more comparable heavy masses in the loops and no GIM mechanism 
anticipated, the LFV branching ratios can be much larger \cite{BorzMas}.

In the leading log approximation the largest contribution comes from the 
left-handed slepton mass matrix yielding a branching ratio
\begin{equation}
      BR21 = \frac{\alpha^3}{G^2_F m^8_s}|(m^2_{LL})_{ji}|^2 \tan^2 \beta,\\
\label{eq:BR21SUSY}
\end{equation}
where
\begin{equation}
     (m^2_{LL})_{ji} = -\frac{1}{8\pi^2}m^2_0 (3+A^2_0/m^2_0)Y^\dagger_{jk}
		\log \left(\frac{M_G}{M_k}\right)Y_{ki}.\\
\label{eq:mLL}
\end{equation}
with the Yukawa couplings specified in the lepton flavor basis and the 
right-handed Majorana matrix diagonal, so $M_k$ is just the kth heavy 
right-handed neutrino mass, while $M_G$ is the GUT scale typically equal 
to $2 \times 10^{16}$ GeV and $m_s$ is some typical SUSY scalar mass.  
Petcov and collaborators \cite{Petcov} have shown that the full evolution 
effects as first calculated by Hisano, Moroi, Tobe, and Yamaguchi \cite{HMTY}
can be extremely well approximated by Eq. (\ref{eq:BR21SUSY}), if one sets 
\begin{equation}
	m^8_s \simeq 0.5 m^2_0 M^2_{1/2} (m^2_0 + 0.6 M^2_{1/2})^2.
\label{eq:m8approx}
\end{equation}

\noindent  We shall see that for the SUSY GUT models to be considered,
the MEG experiment will be able to observe the predicted 
$\mu \rightarrow e\gamma$ branching ratios or place further restrictions
on those models.

\section{Generic Approach to Lepton Flavor Violation in SUSY GUTS}

Many papers have studied predictions for the lepton flavor violating 
processes in SUSY GUT models by adopting a generic approach \cite{generic}.  
Following the procedure of Casas and Ibarra \cite{CasasIbarra}, with the 
charged lepton and right-handed Majorana neutrino mass matrix diagonal, 
the seesaw formula can be inverted to yield the Yukawa neutrino 
coupling matrix
\begin{equation}
  Y_\nu = \frac{1}{v\sin \beta}D_N(\sqrt{M_i})RD_\nu(\sqrt{m_j})
            U^\dagger_{PMNS}\\
\end{equation}

\noindent in terms of a complex orthogonal $R$ matrix which allows for 
various unknown right-handed neutrino mixings.  By using soft SUSY-breaking
benchmarks, adopting various heavy right-handed neutrino masses and 
various $R$ parametric angles and phases, one can ``predict'' the 
radiative LFV branching ratios.  The results are typically presented in 
the form of Monte Carlo scatter plots.  One finds the results strongly
depend on $\tan \beta,\ M_3,\ \theta_{13}$, and the $R$ parameters.
The branching ratios are found to increase by 
powers of 10 as $\theta_{13}$ varies from $0^\circ \rightarrow 10^\circ$,
while the present branching ratio bound for $\mu \rightarrow e\gamma$ 
restricts $M_3$ to lie lower than $10^{14} - 10^{15}$ GeV.

\section{Examples of Predictive SUSY GUT Models}

Instead of adopting a generic approach as described above, we are interested
in determining whether one can differentiate between various SUSY GUT 
models on the basis of their LFV predictions, even if they have similar
predictions for $\sin^2 \theta_{13}$.  For this purpose we have selected
five $SO(10)$ SUSY GUT models which are highly predictive, i.e., their 
model parameters are precisely specified by the authors.

The models differ in the flavor symmetry chosen and the Higgs representations
used to break the $SO(10)$ symmetry at the GUT scale and the electroweak 
symmetry at the weak scale.  We list the models, references, flavor 
symmetry and Higgs representations.
\begin{itemize}
\item AB (Albright-Barr) \cite{AB}: $U(1) \times Z_2 \times Z_2$ with 
         ${\bf 10,\ 16,\ \overline{16},\ 45}$
\item CM (Chen-Mahanthappa) \cite{CM}: $SU(2) \times (Z_2)^3$ with 
         ${\bf 10,\ \overline{126}}$
\item CY (Cai-Yu) \cite{CY}: $S_4$ with ${\bf 10, \overline{126}}$
\item DR (Dermisek-Raby) \cite{DR}: $D_3$ with ${\bf 10,\ 45}$
\item GK (Grimus-Kuhblok) \cite{GK}: $Z_2$ with ${\bf 10,\ 120,\ 
         \overline{126}}$\\[-0.2in]
\end{itemize}
\noindent  Salient features of each model are listed in Table I.  The CM model has a 
relatively large prediction for $\sin^2 \theta_{13}$ which should be easily accessible
to the future Double CHOOZ and Daya Bay reactor experiments.  The AB, CY and DR model
predictions for  $\sin^2 \theta_{13} \sim 0.0025$ are barely within 
reach of those experiments.  The GK model prediction would most likely require a 
neutrino factory to reach that level.  Note that relatively low values of $\tan \beta$
are preferred for four of the models, while the DR model favors a high value. Relatively
mild hierarchies for the massive right-handed neutrinos are predicted for the CM, DR 
and GK models, but the heaviest one occurs in increasing order for these three models,
ranging from $7 \times 10^{12}\ {\rm to}\ 2 \times 10^{15}$ GeV.  The CY model 
predicts a degenerate right-handed spectrum with a lower value of $2.4 \times 10^{12}$.
The AB model, on the other hand, puts the heaviest one at $2.4 \times 10^{14}$ GeV
and the lower two nearly degenerate at $4.5 \times 10^8$ GeV.  Resonant leptogenesis
\cite{leptogenesis} is possibly an interesting feature of that model.  The five 
predictive models studied in this work thus cover a wide range of possibilities and 
suggest that charged lepton flavor violation can play an important role in further 
narrowing the list of viable candidates.

\section{Radiative Lepton Flavor Violation Predictions}

We now turn to the predictions for the branching ratios for $\mu \rightarrow e\gamma$,
$\tau \rightarrow \mu\gamma$ and $\tau \rightarrow e\gamma$ in the five predictive 
models under consideration.  We use the shorthand convention BR21, BR32, and BR31
for these respective branching ratios.  Working in the CMSSM scenario with universal 
soft parameters $m_0,\ M_{1/2}$ and $A_0$ for a given $\tan \beta$ and $sgn(\mu)$, we 
find a variety of plots are possible:

\begin{table}[t]
\begin{tabular}{c|c|c|c|c|c|l}
\hline
Models & Higgs Content & Flavor Symmetry & $M_R$ (GeV) & $\tan \beta$ & 
     $\sin^2 \theta_{13}$ & \qquad\quad Interesting Features \\ \hline\hline
 AB & ${\bf 10},\ {\bf 16},\ \overline{\bf 16}, {\bf 45}$ & $U(1) \times Z_2 \times 
     Z_2$ & 
     $2.4\times 10^{14}$ & 5 & 0.0020 & Large $M_R$ hierarchy with lightest \\
     &  & & $4.5\times 10^8$ & & ($2.6^\circ$)  & two nearly degenerate leads to \\
     & & & $4.5\times 10^8$ & &  & resonant leptogenesis. \\[0.1in]
 CM & ${\bf 10},\ \overline{\bf 126}$ & $SU(2) \times (Z_2)^3$ &  $7.0\times 10^{12}$ 
     & 10 & 0.013 & Large $M_R$ hierarchy with heaviest  \\
     & & &  $4.5\times 10^9$ & &($6.5^\circ$)  & more than 3 orders of magnitude\\
     & & & $1.1\times 10^7$ & & & below GUT scale; large $\sin^2 \theta_{13}$. \\[0.1in]
 CY & ${\bf 10},\ \overline{\bf 126}$ & $S_4$ & $2.4\times 10^{12}$ & 
     10 & 0.0029 & Degenerate $M_R$ spectrum 4 orders\\
     & & & $2.4\times 10^{12}$ & & ($3.1^\circ$) & of magnitude  below GUT scale. \\
     & & & $2.4\times 10^{12}$ & & & \\[0.1in]
 DR & ${\bf 10},\ {\bf 45}$ & $D_3$ &  $5.8\times 10^{13}$ & 50 &0.0024 & Mild $M_R$ 
     hierarchy almost 3 orders\\
     & & &  $9.3\times 10^{11}$ & & ($2.8^\circ$) &   of magnitude  below GUT scale.\\
     & & & $1.1\times 10^{10}$ & &  &\\[0.1in]
 GK & ${\bf 10},\ {\bf 120},\ \overline{\bf 126}$ & $Z_2$ & $2.0\times 
     10^{15}$ & 10 & 0.00031 & Mild $M_R$ hierarchy just  1 order of\\
     & & &  $4.1\times 10^{14}$ & & ($1.0^\circ$) &  magnitude below GUT scale;\\
     & & & $6.7\times 10^{12}$ & & &rather small $\sin^2 \theta_{13}$.  \\[0.1in]
\hline
\end{tabular}
\caption{\label{Table I} Higgs representations, flavor symmetries, and other 
noteworthy features of the five $SO(10)$ SUSY GUT models considered in 
this work.}
\end{table}

\begin{itemize}
  \item  BRij {\it vs.} $M_{1/2}$ for fixed $A_0 = 0$ and different choices of $m_0$.
  \item  $A_0/m_0$ {\it vs.} $M_{1/2}$ scatterplot with a color scheme to indicate the 
  	   branching ratio ranges.
  \item  Ratio of the branching ratios, BR32/BR21 and BR31/BR21 on log-log plots, where
         for example, 
         \begin{equation}
         \log BR32 = \log BR21 + \log \left|\frac{(Y^\dagger_\nu L Y_\nu)_{32}}
         {(Y^\dagger_\nu L Y_\nu)_{21}}\right|^2 \\
         \label{eq:brratio}
         \end{equation}
         \noindent with $L \equiv \log(M^2_\Lambda/M^2_R)$.  Due to the factorization of
         the soft-breaking parameters and the GUT model parameters in the approximation
         of Eqs. (\ref{eq:mLL}) and (\ref{eq:m8approx}), the above yields a straight 
         line with unit slope and intercept given by the second term on the right.  The 
         length of the straight line depends on the soft parameter constraints applied.
\end{itemize}

\noindent  For the sake of brevity, we shall present only the third type of plots.  For 
a more complete description we refer the reader to a longer paper submitted for 
publication \cite{ACII}.

We have imposed the following soft parameter constraints \cite{PDG}:
\begin{equation}
\begin{array}{lrrcr}
	{\rm For}\ \tan \beta = 5, 10: &  m_0:  &   50 &\rightarrow& 400\ {\rm GeV}\\
			  &  M_{1/2}: & 200 &\rightarrow& 1000\ {\rm GeV}\\
			  &  A_0:  &  -4000 &\rightarrow& 4000\ {\rm GeV}\\
	{\rm For}\ \tan \beta = 50: &  m_0:  &  500 &\rightarrow& 4000\ {\rm GeV}\\
	                 & M_{1/2}:  & 200 &\rightarrow& 1500\ {\rm GeV}\\
		       &  A_0:  &  -50 &\rightarrow& 50\ {\rm TeV}\\        
\end{array}
\label{eq:softlimits}
\end{equation}
In addition it is desirable to impose WMAP dark matter constraints \cite{WMAP} in the 
neutralino, stau or stop coannihilation regions \cite{coannih}, where the lightest 
neutralino is the LSP.  These more restrictive constraints are well 
described by the quadratic polynomial for the soft scalar mass in terms 
of the soft gaugino mass \cite{WMAPconstraints}:
\begin{equation}
\begin{array}{rcl}
	m_0 &=& c_0 + c_1 M_{1/2} + c_2 M^2_{1/2},\\
	{\rm where}\ c_i &=& c_i(A_0,\ \tan \beta,\ sgn(\mu)).\\
\end{array}
\label{eq:DMlimits}
\end{equation}
where $m_0$ is bounded since $M_{1/2}$ is bounded.  If $M_{1/2}$ is too 
small, the present experimental bound on the Higgs mass of $m_h \geq 114$
GeV may be violated or the neutralino relic density in the early universe will be too
small, while if $M_{1/2}$ is too large the neutralino relic density
will be too large.  

\begin{figure}[p]
\includegraphics*[scale=0.56]{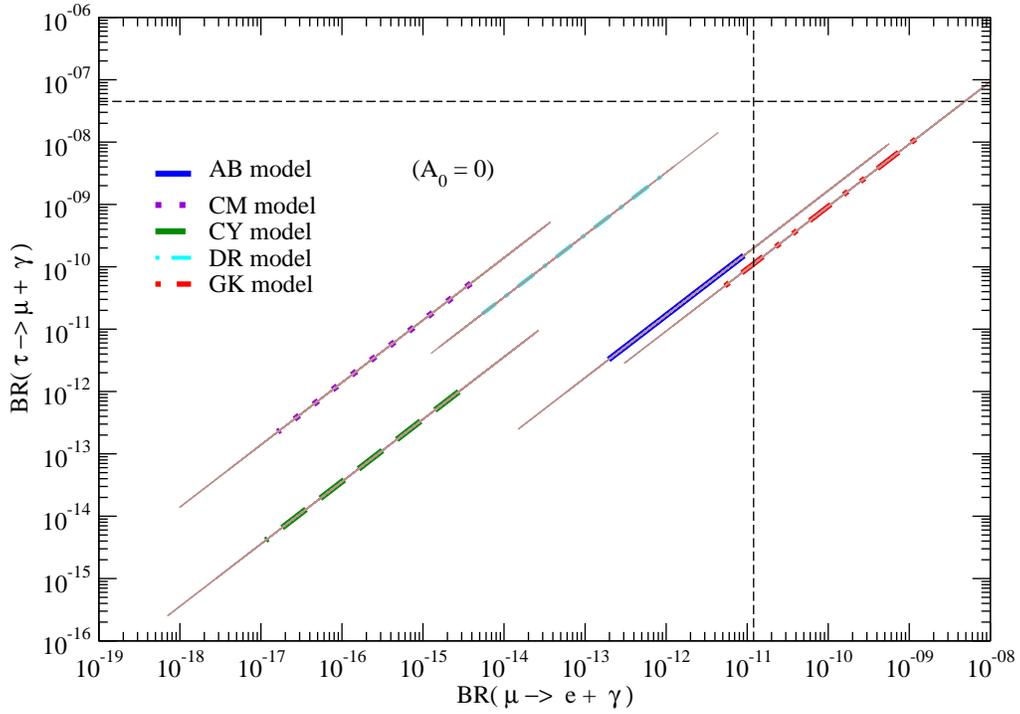}\\[-0.4in]
\caption{Branching ratio predictions for $\tau \rightarrow \mu + 
 \gamma$ {\it vs.} branching ratio predictions for $\mu \rightarrow e + 
 \gamma$ in the five models considered.  The soft SUSY breaking constraints
 imposed apply for the thin line segments, while the more restrictive WMAP 
 dark matter constraints apply for the thick line segments.   The present 
 experimental constraints are indicated by the dashed lines.}
 \end{figure}

\begin{figure}[p]
\includegraphics*[scale=0.56]{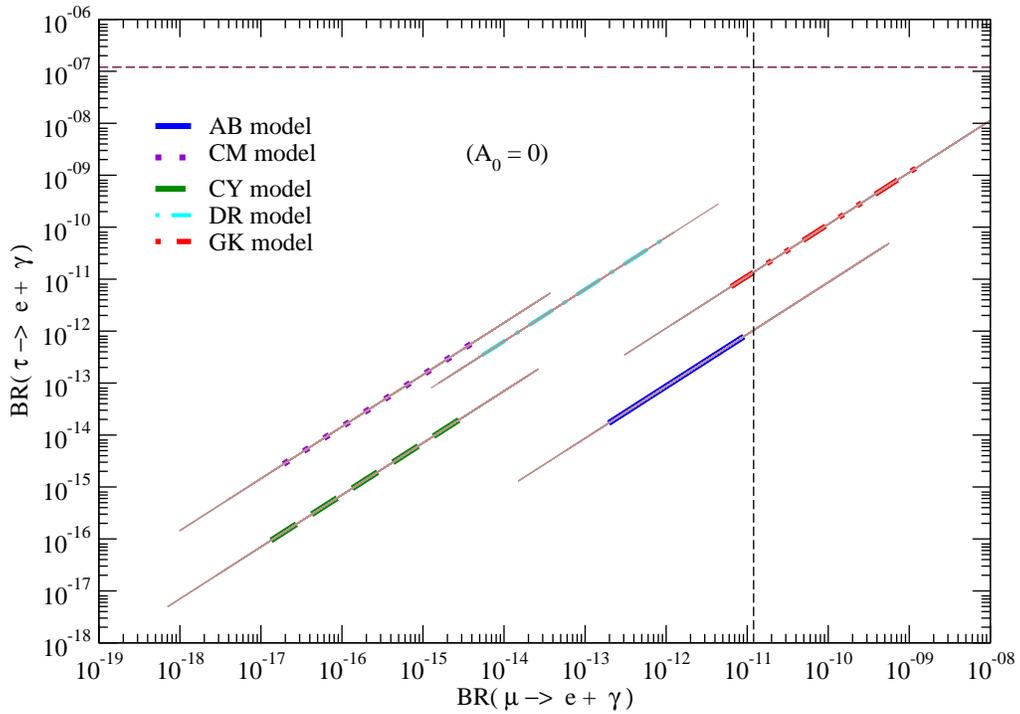}\\[-0.4in]
 \caption{Branching ratio predictions for $\tau \rightarrow e + 
 \gamma$ {\it vs.} branching ratio predictions for $\mu \rightarrow e + 
 \gamma$ in the five models considered.  The same additional conventions
 apply as in Fig. 4,}
 \end{figure}

\newpage
In Figs. 4 and 5 we have plotted BR32 and BR31 {\it vs.} BR21 on log-log graphs.
The thin line segments for each model observe the soft parameters constraints 
imposed, while the heavier line segments observe the more restrictive WMAP dark 
matter constraints.  The vertical dashed line reflects the present BR21 bound 
\cite{MEGA}, while the horizontal dashed line refers to the present BR32 or BR31 
experimental limit, respectively \cite{BRlimits}.  It is clear from these two plots 
that the ongoing MEG experiment stands the best chance of confirming the predictions 
for or eliminating the GK and AB models.  Even with a super-B factory, the 
present experimental bounds on the BR32 and BR31 branching ratios can 
only be lowered by one or two orders of magnitude at most \cite{superB}.

The above figures apply in the case where $A_0 = 0$ is selected for the common
trilinear scalar coupling.  We show in \cite{ACII} that as $|A_0|$ is allowed to depart
from zero, the predicted branching ratios increase.  Hence the line segments in 
Figs. 4 and 5 represent lower limits and extend upward somewhat at $45^\circ$ as
$|A_0|$ is increased.

\begin{figure}[b]
\includegraphics*[scale=0.8]{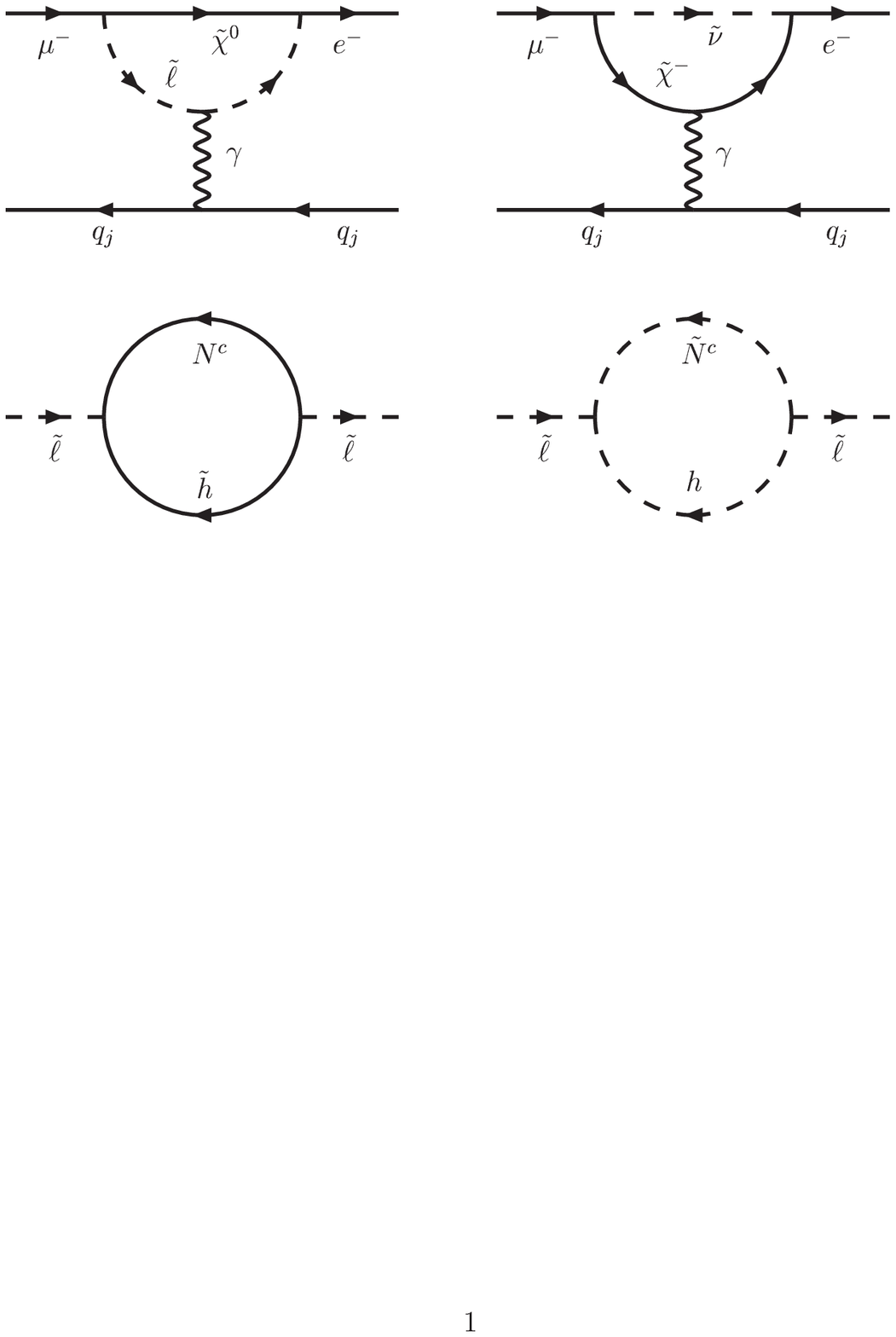}\\[-0.2in]
\caption{Examples of Feynman diagrams for slepton - neutralino 
and sneutrino - chargino contributions to $\mu - e$ conversion in SUSY models
with slepton mass insertions.}
\end{figure}

\section{Lepton Flavor Violation in $\mu -e$ Conversion}

Lepton flavor violation can also occur in the $\mu - e$ conversion process in $^{81}Ti$,
where $\mu + Ti \rightarrow e + Ti$.  The $\mu - e$ conversion branching ratio is 
the conversion rate scaled by the capture rate for the process, $\mu + Ti \rightarrow 
\nu_\mu + Sc$.  The one-loop diagrams involving $\gamma,\ Z$ and Higgs penguins all 
contribute along with box diagrams, but in the CMSSM scenario the $\gamma$ penguin has 
been shown to dominate \cite{AHT}.  We show two such diagrams involving 
slepton-neutralino and sneutrino-chargino loops in Fig. 6, where the effects of 
the virtual $N^c_L$ and $\tilde{N}^c_L$ with their Yukawa couplings appear in slepton
loops.

The $\mu - e$ conversion branching ratio on Ti {\it vs.} BR21 is plotted in Fig. 7 
for the five GUT models, where the tighter WMAP dark matter constraints have been 
imposed, again for the case of $A_0 = 0$. The present conversion branching ratio 
limit for the Ti experiment \cite{mueconv} is shown at $4 \times 10^{-12}$.  It is 
projected that such an experiment will be able to reach down to $10^{-17}$ in a first 
round and down to $10^{-18}$ in a second generation experiment.  While the expectation 
is that the MEG experiment may be able to reach a limit of $10^{-13}$ branching 
ratio for $\mu \rightarrow e\gamma$, it is apparent that a $\mu - e$ conversion 
experiment which eventually reaches a limit of $10^{-18}$ will be considerably 
more powerful.  Potentially such an experiment would be able to eliminate
all five models, and probably all such SUSY GUT models in the CMSSM scenario,
if no positive signal is found.

\begin{figure}[h]
\includegraphics*[scale=0.6]{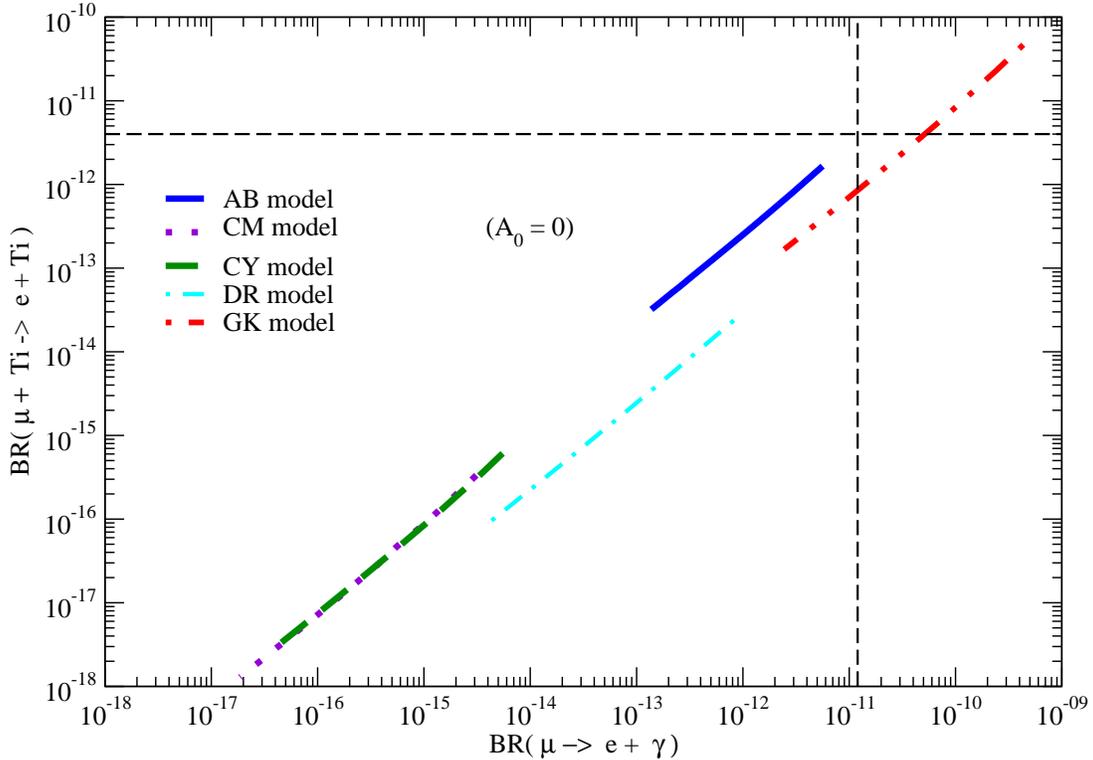}\\[-0.1in]
 \caption{Branching ratio predictions for $\mu - e$ conversion
 {\it vs.} branching ratio predictions for $\mu \rightarrow e + 
 \gamma$ in the five models considered.  The more restrictive WMAP 
 dark matter constraints apply for the thick line segments shown.
 Note that the predictions for the CM and CY models nearly overlap.}
 \end{figure}

\section{Conclusions}

We have tried to differentiate models based on $\sin^2 \theta_{13}$ and charged 
lepton flavor violation predictions.  Our study was initially based on 63 models 
available in the literature prior to June 2006.  There we found that a normal neutrino
mass hierarchy is preferred 3:1.  Moreover, future Double CHOOZ and Daya Bay reactor
experiments will be able to eliminate roughly half of the 63 neutrino models 
surveyed, if their sensitivity reaches $\sin^2 2\theta_{13} \simeq 0.01$ 
as planned.  Still of the order of five models have similar values for 
$\sin^2 \theta_{13}$ in the interval 0.001 - 0.08.  We have suggested that 
charged lepton flavor violation experiments may be able to further distinguish 
them.  If the now-running MEG experiment sees positive signals for 
$\mu \rightarrow e\gamma$, all non-SUSY models or models which do not involve
new physics will be ruled out.  

We then narrowed our study to five predictive $SO(10)$ SUSY GUT models in the 
literature.  All five models have type I seesaw mechanisms implying normal 
hierarchy.  Their predictions for $\sin^2 2\theta_{13}$ lie in the ranges of 0.05 for 
the Chen-Mahanthappa model, 0.01 for the Albright-Barr, Cai-Yu, and Dermisek-Raby
models, and 0.001 for the Grimus-Kubock model.  Previous studies of generic SO(10)
models have concluded that the LFV branching ratios depend critically on $\theta_{13}$
and the heaviest right-handed neutrino mass, $M_3$.  Here we find that $M_3$ 
appears to be more important.  If the MEG experiment can reach an upper bound of 
$10^{-13}$ for the $\mu \rightarrow e\gamma$ branching ratio, it will rule out 
the Grimus-Kubock and Albright-Barr models if no positive signal is seen. 
If a $\mu -e$ conversion experiment can be performed and reach a branching ratio
limit of $10^{-18}$ as projected, it can potentially rule out all five models considered.

\begin{theacknowledgments}
The work reported on here was carried out in collaboration with Mu-Chun Chen.
The author thanks the members of the Theory Group at Fermilab for their kind 
hospitality. Fermilab is operated by the Fermi Research Alliance under 
contract No. DE-AC02-07CH11359 with the U.S. Department of Energy.
\end{theacknowledgments}

\end{document}